\documentclass[utf8]{FrontiersinHarvard} % for articles in journals using the Harvard Referencing Style (Author-Date), for Frontiers Reference Styles by Journal: https://zendesk.frontiersin.org/hc/en-us/articles/360017860337-Frontiers-Reference-Styles-by-Journal
\usepackage[hyperfootnotes=false]{hyperref}
\usepackage{url}
\usepackage{lineno,microtype,subcaption}
\usepackage[onehalfspacing]{setspace}
\usepackage[english]{babel}
\usepackage[normalem]{ulem}

\def\firstAuthorLast{Pfalzner {et~al.}} %use et al only if is more than 1 author
\def\Authors{Susanne Pfalzner\,$^{1,*}$, Stephan Hachinger\,$^{2}$, Jolanta Zjupa\,$^{1}$, Salvatore Cielo\,$^{2}$, Frank W. Wagner\,$^{1}$, Marcus Brüggen\,$^{3}$ and Annika Hagemeier\,$^{1}$}

\def\Address{$^{1}$J\"ulich Supercomputing Centre, Forschungszezentrum J\"ulich, J\"ulich, Germany\\
$^{2}$Leibniz Supercomputing Centre of the Bavarian Academy of Sciences, Garching near Munich, Germany \\
$^{3}$Hamburg Observatory, University of Hamburg, Hamburg, Germany}

\def\corrAuthor{Susanne Pfalzner}
\def\corrEmail{s.pfalzner@fz-juelich.de}

\begin{document}
\onecolumn
\firstpage{1}

\title[Towards FAIR Astrophysical Simulations]{Towards FAIR Astrophysical Simulations} 

\author[\firstAuthorLast]{\Authors} %This field will be automatically populated
\address{} %This field will be automatically populated
\correspondance{} %This field will be automatically populated
\extraAuth{}

\maketitle

% Keywords - if you don't want any simply remove all the text between the curly brackets
%\newcommand{\keywordname}{Keywords} % Defines the keywords heading name

\begin{abstract} \noindent
Reproducibility is a cornerstone of science. FAIR (findable, accessible, interoperable, and reusable) data is often a vital step towards testing the reproducibility of results. The implementation of FAIR principles in the astrophysical simulation community is still varied. We approach the discussion of this topic mainly from a high-performance computing (HPC) point of view. We identify the main obstacles to FAIR astrophysics simulations: First, the vast datasets created in simulations on HPC facilities complicate FAIR data management. Second, missing incentives to fully share codes, results, and diagnostic data. Third, a lack of workflows that include data publication and technical support. Therefore, particularly smaller research groups struggle due to the unavailability of dedicated personnel and time in their efforts towards FAIR and open simulations. We propose actionable steps towards achieving ``FAIRer'' data and open source publication standards in numerical astrophysics.  Our suggestions include low-threshold methods to fulfil the basic FAIR requirements as well as basic tools for FAIR (meta-)data generation and data/code publication. This work is a high-level overview intended to initiate discussions within the community, offering initial solutions to these challenges.
\end{abstract}

{\bf Keywords:} Computational astrophysics, FAIR data, metadata strategies

%\end{frontmatter}

%\tableofcontents

%% \linenumbers

%% main text

\section{Introduction} \label{sec:introduction}

In theoretical astrophysics, the research methodology has evolved significantly due to advances in computing technology. Numerical simulations have become a crucial third pillar alongside observations and analytical theory \citep{Hacking:1983}. In some areas, they have almost replaced analytical methods \citep{Anderl:2015} as they overcome some limitations of analytical calculations, such as the need to formalise solutions for complex differential equations. Computational modelling has enabled tremendous advances in understanding otherwise inaccessible scientific questions and in testing basic theory against observations. In astrophysics, simulations have proven exceptionally relevant as they enable understanding the connection between the evolution of astrophysical objects and the observed snapshots in time. Computational astrophysics is thus an integral link between observations and basic theoretical concepts.

\begin{figure*}[ht]
    \centering
    \includegraphics[width=0.80\textwidth]{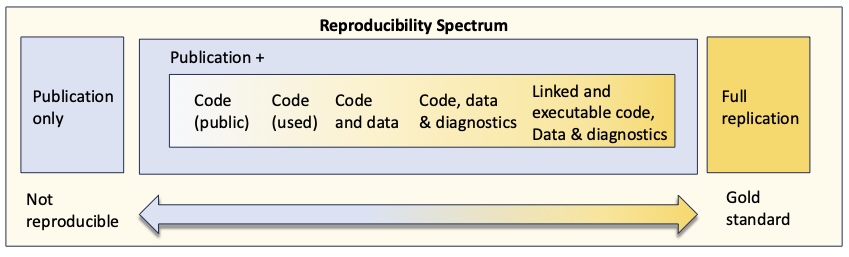}
    \caption{Steps from a ``publication-only'' workflow to one that enables ``full replication'' by providing executable code, data, and diagnostics in a linked format, inspired by the gold-standard illustration for simulation publications by \citet{Peng:2011}. ``Code (public)" refers to the situation where some version of code is publicly available, which is not necessarily the code version used for the specific publication. By contrast, ``Code (used)" refers to the situation in which the exact code used to produce the simulation results is made publicly available. ``Diagnostics" refer to the full set of analysis scripts used to obtain the figures and results presented in the published paper.} \label{fig:gold_standard}
\end{figure*}

Reproducibility is a foundational principle in science \citep{Peng:2011}, as it allows researchers to verify and assess the validity of a model. The FAIR principles -- designed to ensure that data are Findable, Accessible, Interoperable, and Reusable \citep{Wilkinson:2016} -- represent an essential first step towards achieving reproducibility. However, while simulation techniques are advancing rapidly, the reproducibility aspect is often considered unimportant or even neglected \citep{Reinecke:2022}. Publications tend to focus solely on describing numerical methods, but often omit access to essential components such as the codes or the resulting datasets. To achieve true reproducibility, it is essential to provide comprehensive digital research products (DRPs) that include the actual codes, the necessary analysis tools, and all the data involved or resulting \citep{Peng:2011,Samuel:2021,Hassan:2025}. As illustrated in Fig.~\ref{fig:gold_standard}, the ability to reproduce findings is heavily dependent on the availability of complete and transparent information. Furthermore, limited computing resources, hardware variations, and differences in software libraries can complicate reproducibility. In particular, in high-performance computing (HPC) environments, the problem is exacerbated by the generation of extensive datasets. Therefore, promoting the principles of FAIR data is vital to advancing reproducibility in astrophysics.

This work outlines steps for improving the FAIRness and reproducibility of astrophysical simulation data. We identify the necessary data types and reproducibility requirements (Sec.~\ref{sec:reqfair}), review current research data management (RDM) practices (Sec.~\ref{sec:rdm}), and discuss existing RDM tools and methods (Sec.~\ref{sec:tools}). Finally, we examine the gap between requirements and current practices, explore the reasons for the lack of adoption of FAIR methods, and propose next steps to enhance the FAIRness of simulation results (Sec.~\ref{sec:next}).

\section{Requirements for FAIR Simulations and their Reproducibility} \label{sec:reqfair}

First, we analyse the basic requirements to achieve FAIRness and reproducibility in computational astrophysics.

\subsection{From FAIR to reproducible simulations} \label{sec:reqfair-general}

\begin{figure*}[ht]
    \centering
    \includegraphics[width=0.80\textwidth]{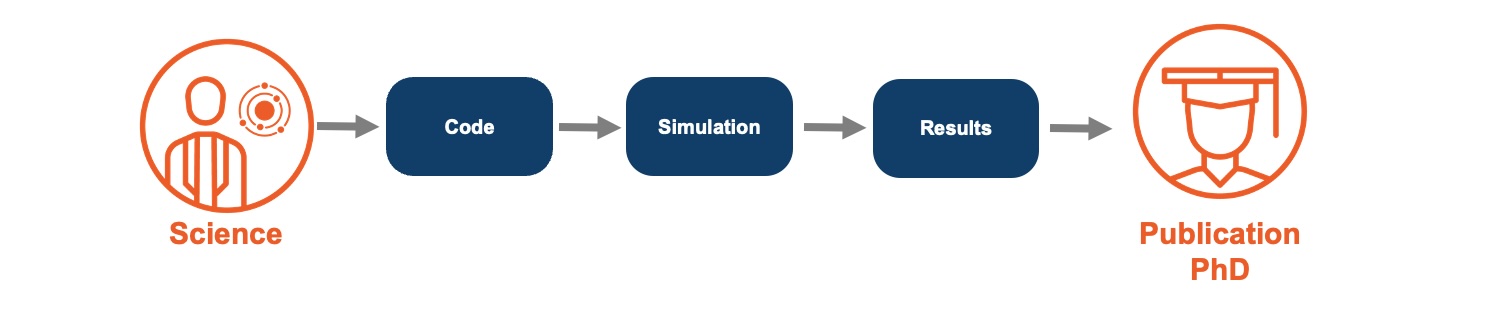}
    \caption{Current workflow from initial scientific idea to publication. The code, simulation setup, and results in full detail are often only documented in the PhD theses, while journal publications provide a more concise, result-focused and therefore often incomplete version.}
    \label{fig:current_workflow}
\end{figure*}

\begin{figure*}[b]
    \centering
    \includegraphics[width=0.80\textwidth]{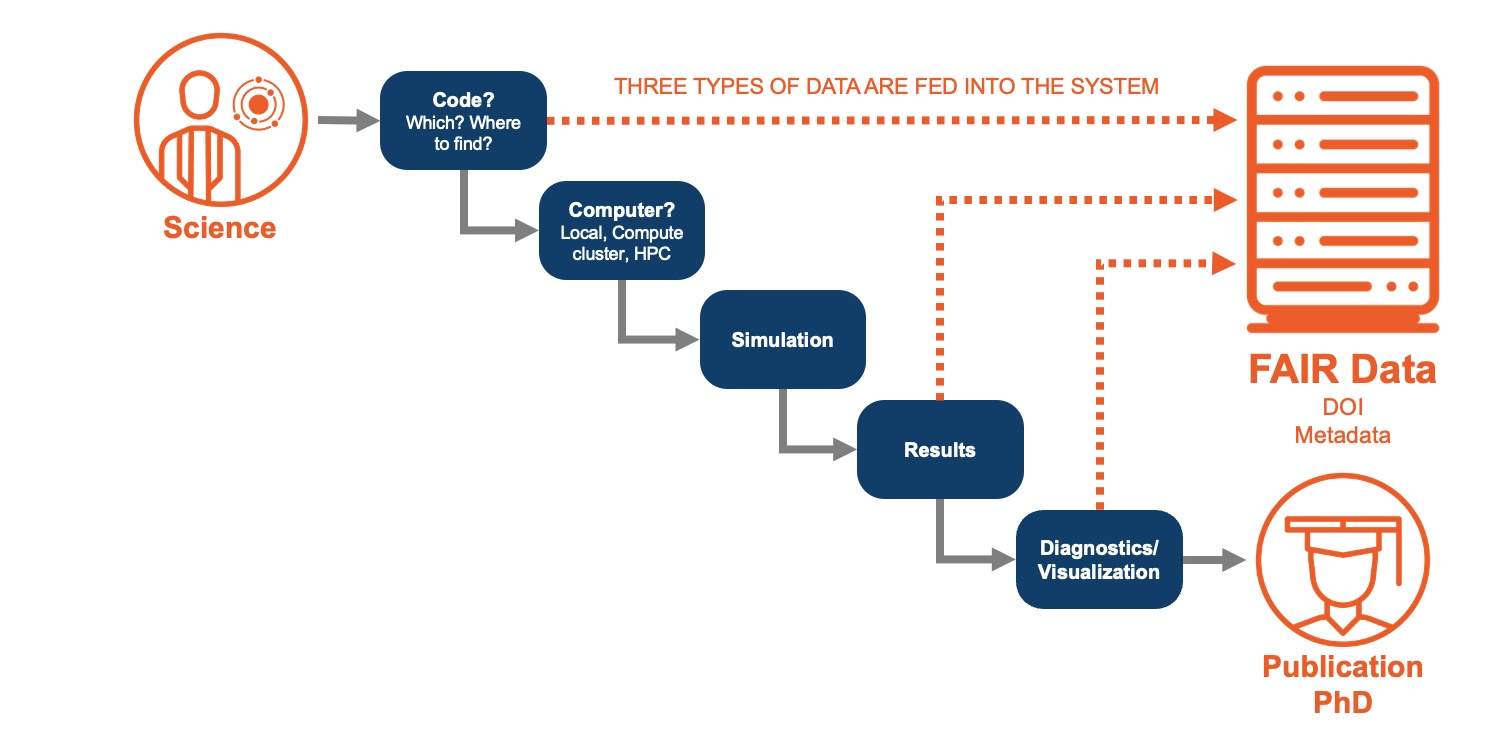}
    \caption{Workflow including the essential components towards a more FAIR-compliant process for astrophysics simulations.}
    \label{fig:fair_workflow}
\end{figure*}

Fig.~\ref{fig:current_workflow} shows a typical simulation workflow. Researchers start with a scientific question, select the appropriate code and machine, run the simulation, diagnose the results, and publish the findings. However, ensuring FAIRness and reproducibility requires more. All components -- the unabridged version of the code used, code-internal modules and parameter choices for a given simulation, (raw) output data, and diagnostic results  -- must be stored and findable for others. Ideally, this is supplemented by information about the software stack, computational environment, and hardware configuration of the used HPC system. The information about the diagnostics should include the analysis scripts that were used to extract published results from the raw output data. This workflow approach is outlined in Fig.~\ref{fig:fair_workflow}. It aims to converge with the ideas of FAIR Digital Objects \citep{Kahn:2006,Schwardmann:2020,Sefton:2025} and PUNCH Dynamic Digital Research Products \citep{the_punch4nfdi_consortium}. To make data findable and reusable, one also needs unique persistent identifiers (PIDs) and sufficient metadata for the data \citep{Wilkinson:2016}. Only by including this full set of components can the data be valuable to others.

Ensuring the reproducibility of simulations extends beyond mere compliance with the FAIR principles \citep{Wilkinson:2016}. The FAIR guidelines do not mandate that data or underlying code be open source, complete, or inherently reproducible. Achieving reproducibility in simulations may require adopting open source code to enable peers and the public to understand scientific results. Additionally, providing open access to results data is essential to enable effective reproducibility testing (open data approach).
We suggest the following additional requirements for making simulation results reproducible:
\begin{enumerate}
 \item Codes and scripts for processing, analysis, and visualisation must be openly available; only for simple codes or well-known methods, it may be sufficient to describe them and comply with
FAIR principles.
 \item Input, intermediate results, and output data (up to the final papers and plots) must be FAIR and open at a sufficient granularity for the reproduction of the computational workflow. 
 \item Workflows, code and data provenance, and the usage of code and data must be sufficiently documented.
\end{enumerate}

These requirements align with an open science approach \citep{UNESCO:2022}. They help to implement the computing-related aspects of open science, focused on open knowledge and infrastructure. In particular, they ensure that researchers extensively document their work, providing essential information in the form of metadata. It is also essential to assign PIDs (e.g., DOIs - ISO 26324:2025, Digital Object Identifier System) to each of the provided data; only then can they be found.

\subsection{Requirements on metadata, identifiers, and provenance} \label{sec:reqfair-components}

For the  workflow and methods sketched above, metadata are equally important as supplying the code, as these metadata explain how the scientist obtained the results. This information is essential for others to reproduce the studies. Ideally, the metadata should clearly and unambiguously specify relevant inclusions and exclusions that allow the community to evaluate the merits independently \citep{Moreau:2024}.

An essential part of the metadata comes with the simulation's input data. The metadata must clarify which choices were made about the code execution. Typical examples of metadata related to the input data include the initial setup and the chosen temporal and spatial resolutions. The metadata must also specify the simulation's workflow. This information could include the number of runs, the intermediate outputs generated, post-processing of the data, diagnostics, and visualisation methods.

A basic requirement of a ``FAIR simulation-based study'' would also be: metadata describing the execution environment(s) for the simulation workflow, as far as this environment can influence the results. The metadata should include information about the computing cluster used and the technical setup (e.g., the number and type of nodes/cores used). This information is critical; for example, the same code run with different precision or different reduction patterns at the end of parallel parts can give different results.

The code provenance should be readily available, which requires a consistent open source repository with a comprehensive commit history. However, basic code provenance tracking can be ensured by an authorship statement, usually included with the licence. To this end, an (open source) licence must require the inclusion of previous licence and authorship/copyright statements in all future versions of the code. Relevant licence assessments are provided by the Open Source Initiative\footnote{\url{https://opensource.org}} (OSI) or the Free Software Foundation\footnote{\url{https://www.fsf.org}} (FSF).

\subsection{Reproducibility} \label{sec:reqfair-reproducibility}

What do we mean by reproducing a simulation result? Repeating the execution of a code to reproduce the result will usually not yield a bitwise identical output, especially on HPC systems. However, scientific reproducibility would require that the execution of the code in a comparable manner lead to identical or comparable output for the same input.

There are several challenges in meeting this requirement. Changes in the hardware and software stack of computing systems can lead to numerically different results \citep{Luo:2024}. Factors outside the control of the researcher conducting the astrophysics simulation -- such as default pinning, interconnect patterns, and the order of operations during reduction (for instance, when summing results from multiple compute nodes) -- can vary between runs, also due to system upgrades. This variability, together with unforeseen factors like bit flips, renders binary-identical reproducibility practically infeasible in astrophysics HPC. Successful reproduction of a simulation would therefore mean recovering the relevant feature in a similar way, possibly on a statistical level. What ``recovering the relevant feature'' means can then be very different from one astrophysics subdomain to another. The definition for the specific case has to be given by scientists deeply entangled in understanding the physical problem.

\section{Current research data management practices} \label{sec:rdm}

In an ideal world, each simulation would fulfil all the requirements listed in Sec.~\ref{sec:reqfair}. Next, we assess the current situation and identify the reasons why this ideal has not yet been achieved.

\subsection{The culture of sharing} \label{sec:rdm-culture}

More and more funding agencies, publishers, and university administrations recognise the importance of a cultural shift towards proper RDM. To enforce this shift, they require that scientists adhere to the FAIR and Open Data principles in their regulations. Fortunately, there is also an understanding that researchers need support to meet these new regulations. For example, the European Open Science Cloud\footnote{\url{https://www.eosc-hub.eu}} (EOSC) has established structures to provide this support \citep{almeida:2017}. Organising a national EOSC node, the ``National Research Data Infrastructure'' (NFDI) initiative in Germany has been built, based on community-built consortia to develop infrastructure frameworks for nationwide RDM. For astrophysics, it is the consortium of Particles, Universe, NuClei, and Hadrons for the NFDI\footnote{\url{https://www.punch4nfdi.de}} (PUNCH4NFDI).

In observational astronomy, RDM standards are widely used, but in astrophysical simulations, Open Data and FAIR practices remain unstandardised.
There exist recommended data and metadata
formats such as those by \citet{Bernyk:2016}, as well as guidelines from major institutions such as NASA\footnote{\url{https://www.nasa.gov/wp-content/uploads/2025/07/nasa-data-strategy.pdf?emrc=488988}} and NSF \footnote{\url{https://www.nsf.gov/funding/opportunities/fairos-findable-accessible-interoperable-reusable-open-science/nsf25-533/solicitation}}. However, these recommendations do not focus on the specific needs of simulation data. The NSF's recent 25-533 guidelines no longer include a discussion of computational reproducibility, as in earlier versions. This may indicate that there is no guiding consensus on the topic yet.

As computational costs increase and reproducibility becomes challenging, a cultural shift in scientific practices is essential, requiring a united effort from researchers, institutions, and funding bodies \citep{Borycz:2023}.

Research on the impact of data sharing \citep{Milham:2018,Perez:2019} found that the lack of recognition and rewards for publishing research data is a barrier to data sharing \citep{Fetcher:2015}. These studies also found that scientists who reported that reusing others' data increased their own efficiency and saved them time are more likely to share their own data \citep{Curty:2017}. The availability of institutional resources supporting data sharing and reuse (e.g., education, technology, expert support) increases the rate of data reuse. Generally, available resources such as repository and metadata creation tools are very beneficial \citep{Kim:2015}.

In astrophysics simulations, adherence to FAIR data and RDM often varies with team size \citep{hagemeier:2023}. Larger teams tend to share codes and results more frequently, whereas smaller teams or individuals may struggle due to limited resources or knowledge gaps. The challenges of sharing data, particularly code, include the need for metadata creation, documentation, and quality control, all of which can be time-consuming.

\subsection{Source code} \label{sec:rdm-code}

The source code of a simulation is the most crucial component when validating, peer reviewing, and reproducing a study. Fortunately, the situation in sharing source code has improved significantly over the last decade. Code is often published on popular repository sites that use Git for version control and collaboration. There are community-specific repositories, such as the Astrophysics Source Code Library\footnote{\url{https://www.ascl.net}} (ASCL), which is a key resource for sharing code in the astrophysics community. The ASCL editors sometimes add source codes without requiring authors to submit them, creating a comprehensive list of codes used in peer-reviewed studies. These codes are indexed by the SAO/NASA Astrophysics Data System \citep[ADS,][]{Kurtz:2000} and the Data Citation Index \citep{Pavlech:2016}, and each code is assigned a unique ASCL number for citation. Users can access the code entry by prefacing the ASCL number with ascl.net (e.g. https://ascl.net/1201.001).

However, there is significant room for improvement in code publishing practices. Best practices would dictate publishing the full source code and documentation along with scientific results. However, it is common to share only a limited version of the code, with essential modules often remaining proprietary, which is delaying reproducibility. Currently, complete code publication is rare, as sharing reduced versions protects developers who invest time in creating new functionality, often in a competitive research environment.

A more technical problem is the practical challenge of compiling old source codes for newer systems. Problems typically arise due to poorly documented build systems, deprecated code constructs, and outdated third-party libraries. The dependency of codes on libraries which again depend on other libraries can lead to a so-called ``dependency hell'', where even the code authors may not exactly understand their reliance on potentially deprecated software.

\subsection{Data: From inputs to post-processed results} \label{sec:rdm-data}

Numerical studies produce a variety of data from code execution, but not all of it needs to be shared. Research adhering to open science principles should make available the data necessary (1) to reproduce the results and (2) for reuse by others. Determining which data meet these criteria is complex and practical guidelines are lacking. Relevant and consistent data products must be shared, involving in particular the input data used to create each output dataset, all metadata necessary to enable reuse, and permanent identifiers for easy referencing.

Technological bottlenecks in RDM are often due to large output data sizes resulting from high dimensionality and high-resolution grids. Simulations often include methods and models for a number of physical processes, such as gravity hydrodynamics, star formation, radiation, and many more, requiring the tracking of numerous properties. Existing solutions mainly focus on uploading datasets to research repositories, but this approach is often impractical due to the size of the datasets. Consequently, facilities for effectively publishing large datasets are lacking, leading to significant amounts of ``dark data'' that may never be used again \citep{Schembera:2020}. Researchers face this challenge not only for the raw data from their code runs but also for their intermediate results, which can be larger than the final results. Intermediate results would be essential for basic data-provenance tracking, as is common in other scientific disciplines.

Astrophysics shares this problem with other computationally intensive sciences, such as Earth-System Sciences \citep[]{EarthCubeRCN2023}. Their data curation faces similar challenges, such as the prohibitive size of simulation outputs, a lack of a standard for model preservation, high computational costs, reproducibility issues, proprietary codes, and undercited software development works. \citet[]{EarthCubeRCN2023} formulated a decision-making rubric with guidelines to estimate what to preserve. Additionally, they distinguish between “knowledge production” and “data production”, whereby “knowledge production" projects typically generate far more output than can be stored in institutional repositories. Astrophysics may want to follow a similar approach in the future.

\subsection{Standards for metadata, formats, and storage systems} \label{sec:rdm-standards}

The International Virtual Observatory Alliance\footnote{\url{https://www.ivoa.net}} (IVOA) and the virtual observatory (VO) following its framework are key organisations providing standards for astrophysical data, primarily focussing on observations and offering compatible object catalogues for tools like Topcat and DS9. The IVOA also supports simulated data through the Simulation Data Access Layer (SimDAL) and Simulation Data Model (SimDM), enabling users to query and retrieve simulated datasets. However, data uploading can be challenging for individual scientists, making these services more appealing to large collaborations. Some simulation portals of such collaborations, such as Magneticum and CosmoSim \citep{Riebe:2013,Klypin:2011,Ragagnin:2017}, offer user-friendly interfaces to the VO site. In the following, we access the metadata-provisioning, format harmonisation, and storage methodologies that a typical astrophysicist performing HPC simulations would use.

\noindent
{\bf Metadata:} Metadata often leave ample room for improvement. If provided at all, they are often not of sufficient quality or do not comply with standards. There are essentially three problems:
\begin{enumerate}
\item Insufficient knowledge on how to provide metadata.
\item Lack of understanding of which metadata are essential. 
\item Perception of metadata supply as an annoying overhead to existing workload.
\end{enumerate}

Researchers play a key role in providing high-quality and standardised metadata (e.g., DataCite \citep{Starr:2011}). They can include metadata in file headers (like in the FITS format) or in separate files, as seen in astrophysics codes such as RAMSES \citep{Teyssier:2002} and ENZO\footnote{\url{https://enzo-project.org}}. The latter simplifies file-type conversions (e.g., BOV, Xdmf). Many formats also support reading from multiple raw files, which eases parallel I/O. Although not essential, having metadata in human-readable forms aids in file searching and content decoding, especially since data are often binary encoded. Avoiding complicated metadata tools can make post-processing easier for users.

\noindent
{\bf Formats:} Unlike in observational astrophysics, there does not exist a standardised data format for simulation data. So far, researchers have resorted to existing data standards such as, for instance,
\begin{itemize}
   \item the well-known \verb|FITS|\footnote{\url{https://fits.gsfc.nasa.gov}} (for 2D or uniform 3D data),
   \item \verb|HDF5|\footnote{\url{https://www.hdfgroup.org/solutions/hdf5}} (very general and can be manipulated via CLI through \verb|hdf5tools| or \verb|h5utils|),
   \item \verb|VTK|\footnote{\url{https://vtk.org}} (preferable for enabling visualisations), and
   \item \verb|AMReX|\footnote{\url{https://amrex-codes.github.io}} (formerly BoxLib, with a large set of tools for grids and particles, also GPU-friendly).
\end{itemize}

However, many of the data do not comply with any of these standards, making reuse very difficult. Influential consortia could promote such formats to promote homogeneity.

\noindent
{\bf Output storage:} Private storage solutions for large input and output data are typically found in centralised computing and HPC data centres. For publishing low-to-medium-sized datasets, Git-based platforms as well as repository solutions from the institute to the international level are often suitable. This is possible due to Git Large File Storage\footnote{\url{https://git-lfs.com}}, which allows Git repositories to manage large files through text pointers while storing the content on remote servers.

\subsection{Data analysis and visualisation scripts} \label{sec:rdm-scripts}

Scripts and pipelines that extract final results from simulation output data are an integral part of most computational studies. A whole subcategory of publications focuses on data analysis and visualisation. Reusing existing data with new or comparative scripts can also boost scientific advances, extracting novel results and insights. Increasingly, such analysis scripts are put under version control systems, such as Git, and made publicly available on platforms like GitHub, GitLab, Bitbucket, or open repositories provided by host academic institutions, or alongside Jupyter notebooks.

What is missing most is to couple those sets of scripts publicly to the published result\textbf{s}. We stress that, for data analysis papers in computational astrophysics, the set of analysis scripts is the core component, and any publication that uses analysis and visualisation scripts on the way to the final result must also share those scripts openly to allow transparency and reproducibility.

\subsection{Current lighthouse solutions for data sharing} \label{sec:rdm-solutions}

Some larger collaborations make their results available in a FAIR manner. These research groups are involved in large-scale simulation projects that share their raw or processed data through dedicated web portals. To access the data, one usually only needs to register. Such lighthouse solutions use storage at supercomputing centres to share the results of their studies. They make data attractive by allowing users to browse data online, perform small-scale analysis, or run data reduction tasks on their HPC systems.

Examples include the Magneticum and CosmoSim web portals \citep{Riebe:2013,Klypin:2011,Ragagnin:2017}. The Virgo consortium has been at the forefront of data-publication activities from cosmological simulations. It has established a landing webpage\footnote{\url{https://virgo.dur.ac.uk/Page_Data/Data/index.html}} for scientific output data from simulation projects conducted within the Virgo framework. These include the Eagle simulations \citep{McAlpine:2016,eagle:2017}.
Another prime example are the seminal cosmological galaxy formation simulation suites Illustris \citep{Nelson:2015} and IllustrisTNG \citep{nelson:2019}, for which the output data is publicly available on website\textbf{s\footnote{\url{https://www.illustris-project.org/data}}}\footnote{\url{https://www.tng-project.org/data}}. Their data publications include output data for a series of snapshots in time, supplemented with various (post-)processed quantities, browsable and downloadable in various ways. In addition, a detailed description ensures reusability. All these portals have proven impactful in fostering scientific work, but require significant setup costs. Nevertheless, there are still points of possible improvement. Metadata are not often standardised, and PIDs are primarily assigned ``indirectly'' as a DOI attached to the publication. The availability of dedicated manpower with capacities, time, funding, and relevant knowledge is the deciding factor that allows large-scale flagship projects to implement FAIR solutions while research groups and individuals running their own small-to-medium-sized simulations often struggle with producing FAIR-compliant, open source data publications.

FAIR practices also vary substantially across subfields:
Outside the computational cosmology and galaxy formation cases detailed above, few noteworthy examples for open data exist, such as the Garching Core-Collapse Supernova Research Archive\footnote{\url{https://wwwmpa.mpa-garching.mpg.de/ccsnarchive/}}, the MHD Turbulence Simulation Archive\footnote{\url{https://www.mhdturbulence.com}}, \texttt{MIST} stellar evolution tracks and isochrones \citep{mistIsochrones2016dotter},
the \texttt{SXS Catalog of merging Black Holes} simulations\footnote{\url{https://data.black-holes.org/simulations/index.html}}, the \texttt{LISA Data Challenges} for gravitational waves \citep{lisadatachallenges2022baghi},
the \texttt{ROCKE-3D Simulations of Planetary Climates}\footnote{\url{https://data.giss.nasa.gov/rocke3d/maps/}}, and the \texttt{CompOSE} tabulated equations of state\footnote{\url{https://compose.obspm.fr/}}. A complete survey of these fast moving fields is however beyond the scope of this article.

\section{Tools for Making Code and Data FAIRer} \label{sec:tools}

The question is how to support smaller teams and individual researchers in making their simulation data FAIRer. They would be educated and enabled to explicitly manage and share data, code, and methods, as it is crucial for enabling other researchers to reproduce the results. A general improvement is already possible if researchers modify their practices \citep{Samuel:2021} by
\begin{itemize}
\item improving record-keeping throughout the research process,
\item using tools for version control for source code and data, ensuring a clear history of the research, and
\item using workflow management systems to organise the steps of a research workflow.
\end{itemize}

Researchers are often unaware of existing tools and guidelines that can help them make their simulation data FAIR. Therefore, we list some existing tools here and also address their limitations. The list is not meant to be complete, but to show useful examples.

\noindent
Open code and data platforms: Several platforms can be used to make the source code of a simulation software available. GitHub\footnote{\url{https://github.com}}, GitLab\footnote{\url{https://about.gitlab.com}}, and Bitbucket\footnote{\url{https://bitbucket.org}} are popular web-based platforms for version control and collaborative software development. Each platform's Git-based version control system allows users to track changes to their code and collaborate with others. 
These platforms also offer features such as issue tracking, project management, and collaboration tools, allowing users to report and track bugs, feature requests, and other issues related to their code.

As mentioned in Sec.~\ref{sec:rdm-code}, the ASCL repository-type platform is a common platform aimed at sharing code within the astrophysics simulation community. The community-focused approach means that the content is more geared towards the needs of potential users.

Apache Subversion\footnote{\url{https://subversion.apache.org}} (SVN) is one of the oldest version control systems that allows users to maintain current and historical versions of files such as source code, web pages, and documentation. Although SVN was a groundbreaking tool in its time, its centralised architecture, limited features, and lack of alignment with modern development practices have made it less popular compared to Git-based platforms like GitHub, GitLab, and Bitbucket. However, SVN is still used in some specific environments where its centralised model is preferred.

Project Jupyter\footnote{\url{https://jupyter.org}} is a web-based platform for interactive computing and data science. Jupyter Notebooks on modern platform architecture offer a range of features to support FAIR code. Researchers are enabled to create code together and share it with other users, and share data and results. Jupyter Notebooks facilitate collaboration and ensure that code changes are thoroughly reviewed and tested. Jupyter's popularity in astrophysics has grown due to its seamless integration with astronomy libraries and tools, such as Astropy and NumPy, making it an ideal platform for data-intensive research.

Zenodo\footnote{\url{https://zenodo.org}} is a general-purpose research data repository. It is primarily known for making research outputs like papers, data, workflows, and software FAIR and citable by assigning DOIs. However, Zenodo also allows users to deposit and share source code. It integrates well with GitHub, allowing users to easily archive and publish versions of a GitHub repository, making the code more discoverable and citable.

\noindent
{\bf Data formats:} There are several tools available for simple data encoding, one of which is the Brick of Values (BOV). This encoding can be particularly helpful during the initial stages of prototyping and code development. When selecting formats for a project, it is essential to choose those compatible with scalable software designed for scientific visualisations, such as VisIt\footnote{\url{https://visit-dav.github.io/visit-website}}, Paraview\footnote{\url{https://www.paraview.org}}, and OSPRay Studio\footnote{\url{https://www.ospray.org/ospray_studio}}.

\noindent
{\bf Code provenance and licencing:} Code provenance and licencing are essential aspects that go together, as licences usually carry authorship notices. The Open Source Initiative\footnote{\url{https://opensource.org}} (OSI) offers valuable assessments of various licences to help ensure that code usage is in accordance with the intended permissions and restrictions. Additionally, the Free Software Foundation\footnote{\url{https://www.fsf.org}} (FSF) provides further guidance and resources on this topic.

\noindent
{\bf Metadata:} Currently, there is a notable absence of dedicated tools for metadata creation specifically tailored for astrophysics simulations, despite the initiatives undertaken by the VO. In contrast, other scientific domains have advanced concepts that facilitate metadata management. For instance, NFDI DataPLANT has implemented the ``Annotated Research Context'', which utilises the internationally recognised ISA (Investigation, Study, Assay) metadata framework \citep{garth:2022}. Unfortunately, a comparable system for simulation codes in astrophysics remains undeveloped, highlighting a gap that could be addressed to enhance data accessibility and usability within the field.

\noindent
{\bf Results:} A growing number of publishers are providing access to the underlying data associated with their published figures. Although this access may not enable a complete reproduction of the results, it facilitates reuse by other researchers. This reuse can manifest itself as a direct comparison with other simulation outcomes or observational data. A notable platform supporting this initiative is Figshare\footnote{\url{https://figshare.com}}, which was developed to promote the principles of FAIR data in research \citep{Bhardwa:2025}.

\noindent
{\bf Containerisation:} Containerisation uses tools such as Docker\footnote{\url{https://www.docker.com}}, Apptainer\footnote{\url{https://apptainer.org}} (formerly Singularity), ReproZip\footnote{\url{https://www.reprozip.org}}, or Podman\footnote{\url{https://podman.io}} to tackle important issues in scientific research, such as reproducibility, portability, collaboration, and scalability. These tools package the entire computing environment above the kernel, including the operating system, software, libraries, and dependencies, into one unit. In this way, workflows can run smoothly on different systems. This consistency makes it easier to share and reuse scientific tools, encourages teamwork, and supports the FAIR principles.

There are some limitations to consider. Although tools like Docker are popular in cloud and development environments, they can face issues in HPC due to security and administrative rules. Apptainer, designed for HPC, offers better compatibility and security by allowing containers to run without root access. 
Although containers can achieve reproducibility in smaller-scale computing use cases (such as a single laptop or workstation), the same cannot be said for supercomputing environments. Containers built on HPC systems cannot avoid interfacing with the host HPC system libraries for efficient distributed-memory parallel computing, notably the filesystem library (such as GPFS) installed on the HPC system or lower-level libraries on which MPI relies (such as PMIx). As HPC system libraries are outside the container scope and are continuously updated, this poses a fundamental limitation on containers providing a framework for (binary-identical) reproducibility of simulation results on HPC systems.

\textbf{\bf Dependency management:} 
Software management based on virtual environments could be a step forward. While not suitable to deliver simulation software itself, tools such as \texttt{spack}\footnote{\url{https://spack.io}} and \texttt{conda}\footnote{\url{https://docs.conda.io/en/latest}} can provide system libraries and dependencies  
even in complex dependency chains. Besides the advantages in automation, standardisation, and software reproducibility, other paramount points are the seamless integration with the module system of HPC machines, the portability of environments across systems and the ability for users to create their own environment. The main limitations of this approach are (i) the availability of pre-built packages (as the difficulty curve for coding own packages can be rather steep), (ii) the limited internet access from some HPC machines for safety reasons, and (iii) licensing limitations and costs of some software repositories (e.g. \url{anaconda.org}) for use on shared systems.

\noindent
{\bf Workflow management:} Computational workflows consist of chained tasks (e.g., data processing and analysis steps) that are bundled into automated pipelines. Workflow management systems orchestrate these tasks with dependency resolution and scaling to address significant challenges in scientific research, such as transparency, reproducibility, and data preservation. Popular options include \texttt{Luigi}\footnote{\url{https://luigi.readthedocs.io/en/stable}}, \texttt{Snakemake}\footnote{\url{https://snakemake.github.io}}, \texttt{Yadage}\footnote{\url{https://yadage.readthedocs.io/en/latest}}, \texttt{REANA}\footnote{\url{https://reana.io}}, \texttt{Nextflow}\footnote{\url{https://www.nextflow.io}}, and \texttt{Airflow}\footnote{\url{https://airflow.apache.org}}. 
Each management system focusses on different functionalities; given the complexity of choosing a suitable system, a comparison of the advantages and disadvantages of various workflow management systems is not easy but necessary \citep{Schmitt:2023}. Notable usage examples include Snakemake orchestrating KM3NeT neutrino telescope simulations in astroparticle physics \citep{Sinopoulou:2025}, the use of Pegasus by the LIGO Scientific Collaboration for the detection of gravitational waves from colliding black holes \citep{Brown:2007}, and AMPEL-Argo being leveraged for real-time multi-messenger astrophysical analysis \citep{Nordin:2019}. Additionally, the PUNCH4NFDI consortium supports REANA for research data management and analysis.

\section{Next Steps Towards FAIR Computational Astrophysics} \label{sec:next}

To ensure that simulations are FAIR and results can be reliably repeated, we need to address both cultural and technological challenges. These challenges often overlap; for example, if tools are not easy to use, researchers may lose motivation. This perceived lack of interest from scientists can slow the creation of new tools. In the following, we provide several strategies to help break this cycle.

\subsection{Motivating a cultural shift} \label{sec:next-cultural_shift}

At the moment, the pressure to improve FAIR standards comes mainly from publishers and funding agencies. 
Nevertheless, FAIR simulations can only be the first step. Simulations should aim not only for FAIR data but also for reproducible simulation results. FAIR data of unreproducible results are worthless.

Currently, the reproducibility of the research results is not adequately tested. This oversight comes from the lack of incentives for researchers to reproduce existing findings. To address this issue, it is essential to enhance the recognition of model-comparison studies and reproduction studies within the scientific community. Such studies can serve as excellent topics for Bachelor’s and Master’s theses. However, to encourage this avenue of research, the current reward system needs to be reformed, as Master's theses that do not yield significant new scientific contributions are often viewed unfavourably as prerequisites for pursuing a PhD.

Today, the number of publications and citations is regarded the strongest indicator of scientific success. However, testing the reproducibility of simulation results would significantly strengthen the credibility of the considered works and accelerate scientific progress. One would need a measure for the reproducibility of a simulation result. A good example\footnote{\url{https://www.acm.org/publications/policies/artifact-review-and-badging-curren}} of how reproducibility can become part of the evaluation process comes from computer science. They developed a badge system with different categories -- artefacts evaluated, artefacts available, and results validated. The lowest category is awarded to papers in which the code and diagnostics are sufficiently documented to be exercisable and include appropriate evidence of verification and validation. An intermediate badge is given to papers in which the associated code, data, and diagnostics have been permanently available for retrieval. This level corresponds to FAIR simulations. The highest badges are reserved for work in which the main results of the paper were obtained in a subsequent study by a person or team other than the authors.

The practice of publishing only a reduced code version would require several changes. The underlying problem is that the significant time investment in developing code is not adequately rewarded.  One solution would be to implement embargoes. An embargo would enable the developers of novel codes to conduct their research with a temporal advantage, ensuring that the development effort pays off. At the end of the embargo, the reproducibility of the results could be tested. Furthermore, one could separate the publication of the full version from that of the astrophysical findings. The code paper should be coupled with an open source version of the code, made publicly available. The separate code paper would be referenced in the astrophysical paper, keeping the latter shorter. Such a practice could also be applied to submodules in larger code projects, as well as to pre-processing and post-processing functionality. This approach ensures that scientists driving the entire scientific process receive credit proportional to their investment. Separate credit for code development allows researchers to build a research software profile according to their strengths.

What makes it especially difficult for astrophysics simulations, compared to other fields, to move to FAIR and reproducible simulations? In astrophysics, often a single scientist or a small group is the driver in developing new simulation schemes. Often, these new methods have a high impact and are later adopted by other communities. An example is hierarchical tree codes, which were initially developed in astrophysics but are now widely used across physics and engineering. Thus, the astrophysics simulation effort focusses on developing new methods rather than on writing a single code to be used by many scientists, as in solid-state physics, climate science, and chemistry. Fields that predominantly use community codes typically devote more resources per code than small groups in astrophysics. Where the user community is larger, the efforts towards FAIR data are also greater in astrophysics simulations. 

Publishers and funding agencies are currently the main drivers towards FAIR data. However, HPC centres can also play a role. HPC centres usually require potential users to demonstrate the performance of the codes. Similarly, HPC centres could require scientists to devise a reproducibility concept before any computing time is granted.

\subsection{Developing supportive tools and workflows} \label{sec:next-support}

Many computational astrophysicists currently do not include essential metadata or assign PIDs to their work. It would be beneficial if all astrophysicists were able to assign DOIs \sout{(ISO 26324:2025, Digital Object Identifier System)} to their data and compile the validated fundamental DataCite metadata required for this process. If including these metadata are not possible, e.g., when dealing with intermediate results that are not yet public -- alternative PIDs, such as B2HANDLE PIDs leveraging handle.net\footnote{\url{https://eudat.eu/service-catalogue/b2handle}} \citep{Kahn:2006}, can be used. These PIDs help identify the datasets clearly and prepare them for FAIR compliance.

\begin{figure*}[ht]
    \centering
    \includegraphics[width=0.75\textwidth]{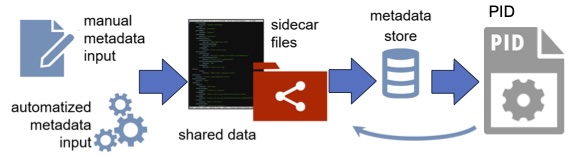}
    \caption{Minimalistic workflow for publishing data directly on high-performance computing (HPC) systems without moving them \citep{Hachinger:2025}, where metadata are stored alongside the data, ingested into a database, exposed via landing pages, and linked with EUDAT-B2HANDLE persistent identifiers (PIDs) and mass-data transfer mechanisms to enable third-party access and reuse.}
    \label{fig:datapub_workflow}
\end{figure*}

In line with the InHPC-DE project, \citet{Hachinger:2025} has suggested a generic procedure to help researchers to make their simulation data FAIRer. As illustrated in Fig.~\ref{fig:datapub_workflow}, it consists of the following elements:
\begin{itemize}
    \item implementing a (file-/database-based) mechanism to store metadata with the data, where the simplest approaches are to store the metadata in ``sidecar files'' with the data, or to use an integrated data-metadata system such as Rucio or iRODS \citep{Barisits:2019,Xu:2017},
    \item enabling the users to feed this mechanism by metadata which are manually or semi-automatically generated,
    \item minting persistent identifiers (PIDs) such as DOIs for the (meta-)data, and
    \item generating web pages with a dataset description and links for actually downloading the data, where the persistent identifiers (PIDs) resolve to their landing pages; this means that users entering, for example, the DOI of a dataset into the doi.org resolver will automatically be directed to the landing page.
\end{itemize}

Various larger and smaller scientific computing centres have devised a similar approach. In InHPC-DE context, the approach has been verified in particular with Earth-System Science datasets. For example, a large meteorological simulation dataset \citep{Bernini:2025} was published pushing metadata (including links for actual data access) from storage to an InvenioRDM metadata catalogue, making the data \citep[see][]{Tartaglione:2024} FAIR.

Adopting these approaches and adapting them to the specific needs of astrophysics simulations would have the advantage of allowing one to profit from synergies between communities. 
However, customisation of the approach to the specifics of the community is needed. Adoption or further development of HPC-ready data and metadata standards beyond DataCite, such as those proposed by the IVOA (see Section \ref{sec:rdm-standards}) or those realised by engineers in Metadata4Ing \citep{Iglezakis:2025}, can be an important step here.

The challenge of old software on new computers will continue due to the rapid evolution of computer hardware. Currently, there are no established standards to ensure that codes and data storage remain future-proof. A possible example of such standards could be derived from the archival methods used by state archives. These archival methods not only focus on ensuring bitwise correctness in storage but also guarantee that data formats remain accessible and readable even after decades. Creating similar methods to preserve the source code will be even more complex. For smaller workloads, containerisation or storing virtual machine images could provide a solution, allowing future re-runs of the operating system, libraries, and code in a sandboxed environment. However, as mentioned in Sec.~\ref{sec:tools}, containerising workloads on HPC systems is known to present challenges, particularly regarding MPI communication and performance \citep{Kumar:2022}.

The usefulness of FAIR data would be greatly enhanced if one would link the different data products (code, results data diagnostics, publication, etc). This integrated approach is envisaged by the PUNCH Digital Research Product\footnote{\url{https://rpr-p4n.aip.de}} (DRP). It relates the data products and the specific software or systems utilised in their generation through appropriate PID references. Combining a FAIR  with cross-identity methods will not only improve data transparency, but will also facilitate better management and utilisation of data resources.

\noindent
{\bf Further progress:} Automating computational workflows and data lifecycles can make the documentation process more appealing. By implementing orchestration strategies, the essential information required for a thorough documentation can be automatically collected. This approach reduces the burden of planning, allowing researchers to focus more on their scientific inquiries rather than the logistics of documentation.

\section{Summary and Conclusions} \label{sec:conclusions}

Funding agencies increasingly recognise the need for proper RDM practices that comply with the FAIR and Open Data principles. 
Although observational astronomy has made strides in adopting these standards, astrophysical simulations have yet to catch up.
We examined what is needed to ensure FAIRness and reproducibility in computational astrophysics simulations.

Being reproducible requires more than just following FAIR standards, which mainly focus on identifying data uniquely, but do not require that the data or code be open to the public. The main requirements for reproducible simulations are the following:
\begin{itemize}
\item Codes and scripts should be openly available with comprehensive documentation.
\item Input, intermediate results, and output data must be accessible and follow the FAIR principles.
\item Documentation of workflows, code provenance, and data usage is essential. 
\end{itemize}

Reproducing simulation results requires that the execution of the code yields identical or statistically similar outputs, even if the hardware or software changes. Therefore, the metadata must inform the input data choices, how the code was executed, and the workflow components.

Data sharing practices vary considerably among different sub-communities. Codes are the most shared data products, while missing interfacing of HPC storage to external data ecosystems can be an obstacle to sharing large output data. A key barrier to adhering to the FAIR principles \textbf{in the absence of easy methods} is the lack of human resources. Thus, especially small groups and single researchers need more support in the form of low-effort tools tailored to their needs. Another hurdle is the lack of recognition for data publication, which slows progress.

To help smaller teams and individual researchers, we recommend improved record-keeping, version control, and the use of workflow management systems. A variety of platforms exist already for code sharing, some support also data visualisation and collaboration with others. However, dedicated tools for metadata creation specific to astrophysics simulations are still lacking.

We find that cultural and technological changes are needed to enhance FAIRness and reproducibility in astrophysical simulations. Improving recognition of model-comparison and reproduction studies is vital, and a shift towards a sharing culture could be aided by reforming the reward systems. Incentives for reproducibility should be improved, with potential badge systems indicating data/code availability and a structured approach to publishing code alongside research findings.

On the technical side, using essential metadata and persistent identifiers (PIDs), such as DOIs, for datasets is crucial to FAIR compliance. Here, a connection between the different PIDs of the various data products of the simulation results needs to be developed. Improving documentation practices among computational astrophysicists is also necessary to advance FAIR principles in the field.

In recent decades, the sharing of code among researchers has increased significantly because scientists realised the advantages they gained from code sharing. Therefore, there is justified hope that the move to FAIR and reproducible simulations will soon become standard practice.

\phantomsection
\section*{Acknowledgments} % The \section*{} command stops section numbering

\addcontentsline{toc}{section}{Acknowledgments} % Adds this section to the table of contents

This work was carried out within the framework of the PUNCH4NFDI consortium supported by DFG fund NFDI 39/1, Germany. Parts of the manuscript have also been supported by the project InHPC-DE, funded by the German Federal Ministry of Education and Research (F\"orderkennzeichen 16HPC02), and the BMBF project 01-1H1-6013 AP6-NRW Anwenderunterstützung SiVeGCS.

SC thanks Dr Matteo Foglieni for the interesting discussions about journals on open source software. SP thanks Dr Bernd Mohr for pointing out the badge system for reproducibility of code results in the HPC context. The authors thank our colleagues from the InHPC-DE project, from HLRS, JSC, and LRZ, for the excellent collaboration on FAIR HPC data in general.

%	REFERENCE LIST

\bibliographystyle{elsarticle-harv}
\bibliography{references}

\phantomsection

\end{document}